\documentclass[sigconf]{acmart}

\usepackage{amsmath}

\usepackage{tikz}
\usepackage{booktabs}
\usepackage{multirow}
\usepackage{makecell}
\usepackage{threeparttable}
\usepackage{tabularx}
\usepackage{array}
\usepackage{colortbl}
\usepackage{caption}
\usepackage{afterpage}

\usepackage{xcolor}
\usepackage[most]{tcolorbox}

\usepackage{enumitem}
\usepackage{setspace}

\usepackage{listings}
\usepackage{algorithmic}
\usepackage[linesnumbered,ruled,vlined]{algorithm2e}

\usepackage{pifont}
\usepackage{xspace}
\usepackage{url}
\usepackage{ifpdf}
\usepackage{stackengine}

\newcommand{\cmark}{{\color{green!60!black}\ding{52}}}

\newcommand{\pmark}{%
{\color{orange!90!black}%
\stackon[-4pt]{%
\ding{52}%
}{%
\hspace{0.2em}%
\tikz{\draw[line width=1pt,line cap=round]
(0,0) -- (0.35em,-0.35em);}%
}%
}}

\definecolor{full}{RGB}{198,239,206}     
\definecolor{partial}{RGB}{255,235,156}  
\definecolor{none}{RGB}{240,240,240}     
\definecolor{RepoBlue}{RGB}{0,86,179}

\newcommand{\thetool}{\mbox{\texttt{AgentGuard}}\xspace}

\newcommand{\repo}{\textcolor{RepoBlue}{\url{https://github.com/WhitzardAgent/AgentGuard}~}}

\def\Snospace~{\S{}}

\setcopyright{none}
\renewcommand\footnotetextcopyrightpermission[1]{}

\begin{document}

\title[AgentGuard]{AgentGuard: An Attribute-Based Access Control \\ Framework for Tool-Use LLM-Based Agent}

\author[AgentGuard]{
Jiaqi Luo$^{\dag\ast}$,
Songyang Peng$^{\dag\ast}$,
Jiarun Dai$^{\dag}$\textsuperscript{\S},
Zhile Chen$^{\dag}$,
Zhuoxiang Shen$^{\dag}$,
\\
Geng Hong$^{\dag}$,
Xudong Pan$^{\dag\ddag}$,
Yuan Zhang$^{\dag}$,
Min Yang$^{\dag}$\textsuperscript{\S}
\\[1ex]
$^{\dag}$Fudan University \quad
$^{\ddag}$Shanghai Innovation Institute
\\[0.5ex]
\textsuperscript{\S} Co-corresponding authors
}


\begin{abstract}
LLM-based agents have recently attracted significant attention due to their ability to autonomously invoke relevant tools to accomplish complex tasks.
However, recent studies have shown that these agents face severe security risks, which may lead to privacy leakage, financial loss, or even full system compromise.
In this paper, we present \thetool, an attribute-based access control framework for tool-use LLM-based agents. 
\thetool adopts a client-server architecture. 
On the client side, \thetool provides lightweight integration for agents implemented in different programming languages and architectures.
It requires only minor code modifications (e.g., around 10 lines) without changing the underlying agent execution logic.
On the server side, \thetool provides three complementary inspection mechanisms to cover both single-tool and cross-tool security risks in agent execution.
In addition, it offers a visualized front-end interface for security policy specification and runtime auditing.
Currently, \thetool is publicly accessible at \repo.

\end{abstract}


\maketitle

\begingroup
\renewcommand\thefootnote{}
\footnotetext{* These authors contributed equally.}
\footnotetext{
Email:
\{jqluo24, sypeng23, zhilechen25, zxshen22\}@m.fudan.edu.cn;
\{jrdai, ghong, xdpan, yuanxzhang, m\_yang\}@fudan.edu.cn
}
\endgroup

\section{Introduction}
\label{sec:intro}
LLM-based agents have recently attracted significant attention due to their ability to autonomously invoke relevant tools to accomplish complex tasks~\cite{wang2024survey, xi2025rise}. 
Mainstream agent applications, such as Codex~\cite{codex}, and OpenClaw~\cite{openclaw}, are attracting millions of users.
However, recent studies~\cite{liu2025make,luoautonomy,debenedetti2024agentdojo,zhan2024injecagent} have revealed that such agents face severe security risks, which may ultimately result in privacy leakage, financial loss, or even full system compromise.

\noindent\textbf{Research Gap.}
In response to these threats, a large body of research has sought to improve the security of LLM-based agents~\cite{kim2026sok}. 
Generally, existing approachs fall into two categories: \textit{model-level} defenses and \textit{system-level} defenses.
\ding{182} At the model level, prior approaches~\cite{chen2025secalign,liu2025datasentinel,jiang2025think} attempt to enhance the robustness of the underlying models in agents through prompt engineering and fine-tuning. 
However, recent studies~\cite{zhan2025adaptive,wang2024badagent} have shown that defenses relying solely on improving model capabilities can still be bypassed by adaptive attacks.
\ding{183} At the system level, some works~\cite{wu2024isolategpt, wang2025agentspec, li2025ace} leverage auxiliary LLMs to detect and intercept malicious intentions during agent execution.
Other approaches~\cite{ji2026taming,shi2025progent,debenedetti2025defeating} attempt to ensure security by enforcing predefined security policies. 
Nevertheless, as summarized in \autoref{tab:comparison}, these solutions still suffer from insufficient risk coverage, poor compatibility with existing agents, and complex policy specification.

\noindent\textbf{Our Work.}
To bridge these gaps, we present \textbf{\thetool}, a mandatory access control framework for LLM-based agents.
In general, \thetool adopts a \textit{client-server architecture}, where the two components communicate through a network.
\ding{182} On the client side, \thetool provides lightweight integration for agents implemented in different programming languages and architectures. 
It requires only minor code modifications (e.g., around 10 lines) without changing the underlying agent execution logic. 
During runtime, the \thetool client monitors each tool invocation, forwards the relevant contextual information to the server, and enforces the server’s decision to either allow or deny the action.
\ding{183} On the server side, \thetool provides a visualized front-end interface that allows users to flexibly configure security policies through intuitive parameterized forms. 
During runtime, the server receives tool-related information from the client and performs inspection through three complementary mechanisms, i.e., rule-based detection, LLM-assisted detection, and manual verification, to cover both single-tool and cross-tool security risks in agent execution.

\begin{figure}[!t]
    \centering 
    \includegraphics[width=.45\textwidth]{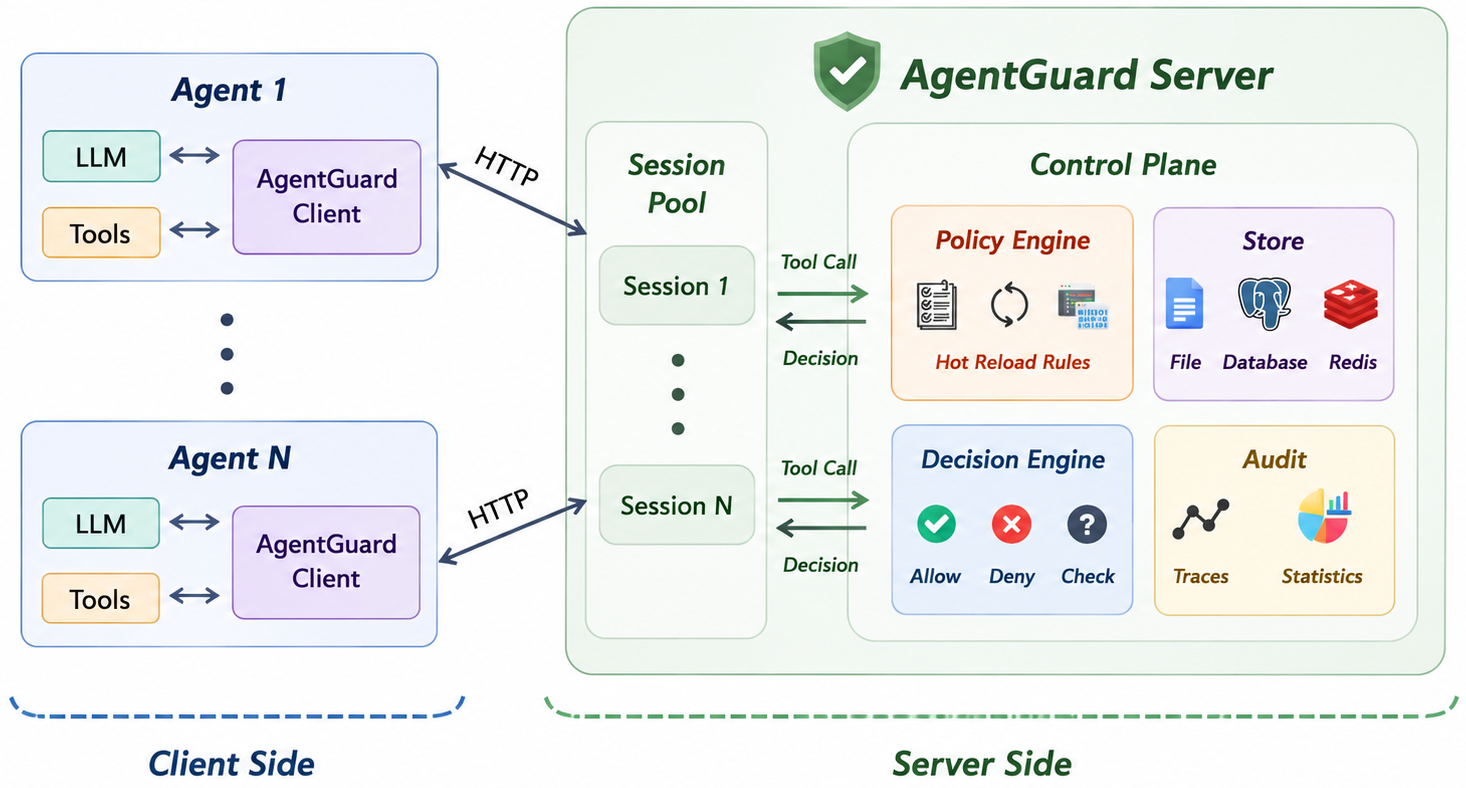}
    \caption{Architecture of \thetool.}
    \label{fig:overview}
\end{figure}

\begin{table*}[t]
\centering
\caption{Comparison with existing works.}
\label{tab:comparison}
\footnotesize
\setlength{\tabcolsep}{3.5pt}
\renewcommand{\arraystretch}{1.3}

\renewcommand\cellalign{cc}

\begin{threeparttable}

\begin{tabular}{lccccccccccccc}
\toprule
\multirow{3}{*}{\textbf{Approach}} 
& \multicolumn{2}{c}{\textbf{Risk Type}} 
& \multicolumn{2}{c}{\textbf{Intervention Phase}}
& \multicolumn{3}{c}{\textbf{Decision Mechanism}} 
& \multicolumn{2}{c}{\textbf{Compatibility}} 
& \multicolumn{4}{c}{\textbf{Usability}} \\
\cmidrule(lr{.5em}){2-3} 
\cmidrule(lr{.5em}){4-5} 
\cmidrule(lr{.5em}){6-8} 
\cmidrule(lr{.5em}){9-10} 
\cmidrule(lr{.5em}){11-14}
& \makecell{Single-\\tool} 
& \makecell{Cross-\\tool} 
& \makecell{Pre-\\execution} 
& \makecell{Post-\\execution} 
& \makecell{Rule-\\based} 
& \makecell{LLM-\\based} 
& \makecell{Manual \\ Verify} 
& \makecell{Framework-\\agnostic} 
& \makecell{Language-\\agnostic} 
& \makecell{Agent-\\decoupled} 
& \makecell{Low-\\intrusion} 
& \makecell{Visual Rule \\ Setup \& Audit} 
& \makecell{Dynamic\\ Rule Update} \\
\midrule

AgentDoG~\cite{agentdog}
& \cmark & \cmark
& \cmark & 
&  & \pmark\tnote{1} & 
& \cmark &  & 
&  &  &  \\

ClawSentry~\cite{clawsentry}
& \cmark & \pmark\tnote{2}
& \cmark &  \pmark\tnote{3}
& \pmark\tnote{2} & \pmark\tnote{1} & \cmark
& \cmark & \cmark & \cmark
& \cmark &  &  \\

AgentBound~\cite{buhler2026agentbound}
& \cmark & 
& \cmark & 
& \cmark &  & 
&  &  & 
&  &  &  \\

CaMeL~\cite{debenedetti2025defeating}
& \cmark & \cmark
& \cmark & 
& \cmark &  & 
& \cmark &  & 
& \cmark &  &  \\

SECAgent~\cite{ji2026taming}
& \cmark & \cmark
& \cmark & 
& \cmark &  &
& \cmark &  
&  &  &  &  \\

ACE~\cite{li2025ace}
& \cmark & \cmark
& \cmark & 
&    & \pmark\tnote{1} &
&  &  & 
&  &  &  \\

Agent Gov. Toolkit~\cite{agt}
& \cmark & 
& \cmark & \pmark\tnote{3}
& \cmark &  \pmark\tnote{1} & 
& \cmark & \cmark & \cmark
& \cmark &  & \cmark \\

Progent~\cite{shi2025progent}
& \cmark & \cmark
& \cmark & 
& \cmark &  & 
& \cmark &  & 
& \cmark &  &  \\

AgentSpec~\cite{wang2025agentspec}
& \cmark & 
& \cmark & 
& \pmark\tnote{4} & \pmark\tnote{1} & \cmark
& \cmark &  & 
&  &  &  \\

IsolateGPT~\cite{wu2024isolategpt}
& \cmark & \cmark
& \cmark & 
&    & \pmark\tnote{1} & \pmark\tnote{5}
&  &  & 
&  &  &  \\

\rowcolor{gray!13}
\textbf{\thetool (Ours)} 
& \cmark & \cmark 
& \cmark & \cmark
& \cmark & \cmark & \cmark 
& \cmark & \cmark & \cmark 
& \cmark & \cmark & \cmark \\
\bottomrule
\end{tabular}
\begin{tablenotes}
\footnotesize
\item \cmark~indicates full support, \pmark~indicates partial support, and blank entries indicate no support.

\item[1] Existing LLM-based verification approaches rely on a unified prompt, preventing users from customizing prompts for specific scenarios or security requirements.

\item[2] ClawSentry cannot effectively extend security policies for multi-tool scenarios. Moreover, when a specific multi-tool execution flow is detected, it relies solely on LLM-based judgment to determine whether the action should be blocked, rather than enforcing user-defined security policies.

\item[3] In both ClawSentry and Agent Gov. Toolkit, post-execution rules are hard-coded. Supporting new rules requires manual code implementation.

\item[4] AgentSpec allows users to define custom security policies, but it requires substantial amounts of manual code implementation of rule-specific detection logic.

\item[5] IsolateGPT only triggers human review at the LLM’s request, preventing user-defined verification conditions.
\end{tablenotes}
\end{threeparttable}
\end{table*}

\noindent\textbf{Contributions.}
Our contributions are summarized as follows:

\begin{itemize}[noitemsep, leftmargin=*]

\item We propose \thetool, an attribute-based access control for tool-use LLM-based agent. 
\thetool adopts a lightweight client-server architecture that can be seamlessly integrated into existing agents, without modifying underlying logic.

\item We develop the \thetool as a centralized security service with audit-oriented observability. 
The server can manage and inspect multiple concurrent agent sessions, while supporting visualized policy specification, runtime monitoring, and security auditing.

\item To support future research and community development on secure LLM-based agents, we publicly release \thetool~at \repo.

\end{itemize}

\section{\thetool}
\label{sec:approach}


\subsection{Overview}
\label{subsec:overview}

As illustrated in ~\autoref{fig:overview}, \thetool is an attribute-based access control for tool-use LLM-based agent that adopts a lightweight client--server architecture.

\begin{itemize}[noitemsep, leftmargin=*]

\item \textbf{Client.}
\thetool offers SDKs for multiple programming languages and requires only minimal code modifications (i.e., approximately 10 lines) to integrate with existing agent frameworks without altering the underlying execution logic. 
During runtime, the client monitors each tool invocation, forwards the relevant contextual information to the server, and enforces the server’s decision to either permit or deny the requested action.

\item \textbf{Server.}
\thetool provides a visualized front-end interface that enables users to flexibly configure security policies through intuitive parameterized forms. During runtime, the server receives information from the client and performs security inspection through three complementary mechanisms, namely rule-based detection, LLM-assisted detection, and manual verification. 
\thetool can address both single-tool and cross-tool security risks during agent execution.

\end{itemize}

As summarized in ~\autoref{tab:comparison}, compared with existing defense approaches, \thetool provides broader coverage of security risks, supports more flexible decision-making mechanisms, and can be seamlessly integrated into existing agent frameworks across different programming languages and architectures. Furthermore, \thetool offers a visualized front-end interface that improves usability and simplifies security policy management.

\subsection{Usage}
\label{subsec:usage}
Users can complete the deployment of \thetool and start using it in just four steps: launching the control server,  performing agent-side setup, configuring access control policies, and monitoring agents in runtime through the visualized front-end.

\textbf{Launch of the Control Server.} The control server is the central component of \thetool. It receives runtime events from agents, evaluates them with the active policies, and returns access control decisions. This separates security control from the agent's own planning and task logic. It also allows multiple agents on different machines to be managed under one policy view.

\textbf{Agent-side Setup.} On the agent side, \thetool is inserted between the agent framework and its tools. The agent provides principal information such as identity, session, role, and trust level. During execution, each tool call is checked by the control server before the final action is taken. Adapters are provided for several mainstream agent frameworks, such as LangChain, AutoGen, and OpenAI Agents SDK, allowing users to integrate \thetool with minimal code and without modifying framework internals or heavily refactoring existing agents.

\textbf{Configuration of Policies.} \thetool uses declarative policies to describe when a tool call should be allowed, denied, or sent for review. A policy can refer to the agent identity, tool attributes, tool arguments, target addresses, and session history. This allows both simple checks on a single tool call and broader checks over a sequence of actions. Policies can be written in the DSL or configured through the visualized front-end.

\textbf{Runtime Monitoring via Visualized Front-end.} The visualized front-end gives users a runtime view of connected agents, their tools, recent activity, and pending review requests. For each relevant tool call, users can inspect the matched rules, final decision, and audit metadata. This helps explain why an operation was allowed, denied, or escalated. The same interface can also be used to configure policies and refine them based on audit records.

\subsection{Access Conditions}
\label{access}
\thetool has been publicly released at \repo under the MIT License.
It is freely available for academic research, educational purposes, and other non-commercial or commercial applications permitted by the license. 
We kindly request that any use of \thetool in academic publications or derived projects appropriately acknowledge the project and cite our paper when applicable.

\subsection{Future Plan}
\label{subsec:plan}
In future work, we plan to extend \thetool to support more mainstream agent frameworks, enable protection for multi-agent scenarios, and incorporate monitoring of LLM inputs and outputs with automatic security policy recommendations tailored to different agent implementations.
We also warmly welcome contributions from the community to help maintain and improve this framework. 
Please feel free to submit issues and pull requests.
\section{Conclusion}
\label{sec:conclusion}
In this paper, we present \thetool, an attribute-based access control for tool-use LLM-based agent. 
\thetool can be seamlessly integrated into existing agent applications, while supporting the detection and mitigation of a wide range of security risks through multiple complementary auditing mechanisms.
Currently, \thetool can be accessed through the website: \repo.

\bibliographystyle{ACM-Reference-Format}
\bibliography{bib}

\end{document}